\documentclass[10pt,twocolumn,a4paper]{article}
\usepackage[pdftex]{hyperref}
\usepackage[pdftex]{graphics}
\usepackage{graphicx}
\usepackage {layout}
\usepackage{amsmath}
\title{Field structure and electron life times in the MEFISTO Electron Cyclotron Resonance Ion Source}
\author{M. Bodendorfer$^{1,2}$, K. Altwegg$^2$, H. Shea$^{1}$, P. Wurz$^2$ \\ \small{\textbf{1:} EPFL - Ecole Polytechnique Federale de Lausanne, Switzerland;} \\ \small{\textbf{2:} University of Bern, Switzerland} }
\begin{document}

\maketitle

\begin{abstract}
\textbf{The complex magnetic field of the permanent-magnet electron cyclotron resonance (ECR) ion source \textit{MEFISTO} located at the \textit{University of Berne} have been numerically simulated. For the first time the magnetized volume qualified for electron cyclotron resonance at 2.45 GHz and 87.5 mT has been analyzed in highly detailed 3D simulations with unprecedented resolution. New results were obtained from the numerical simulation of 25211 electron trajectories. The evident characteristic ion sputtering trident of hexapole confined ECR sources has been identified with the field and electron trajectory distribution. Furthermore, unexpected long electron trajectory lifetimes were found.}
\end{abstract}

\section{Introduction}
\label{intro}

Unlike the magnetic field of solenoid coils the field of permanent magnets can only be tuned over a limited range after the manufacturing process and the construction of the ECR ion source. Great care has therefore to be taken in the design of the magnetic configuration because empirical adaptation of the magnetic field after construction is limited to very few options. The magnetic arrangement chosen for the MEFISTO \cite{AdrianMarti} ion source features a field distribution which cannot be reduced from three to two dimensions for the purpose of a more economic simulation. Hence appropriate 3D simulations have to be performed to bring out the full spatial extent of the field distribution.

\begin{figure}[h!]
\includegraphics[angle=0, width=0.5\textwidth]{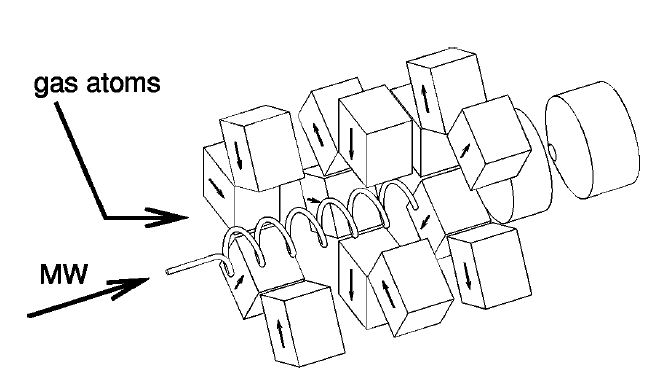}
\caption{\textit{The arrangement of the full permanent magnets of the MEFISTO ECR ion source at the University of Bern (Marti et al, 2001). Some magnets are removed in this drawing for better visualization}}
\label{magnets_MEFISTO}
\end{figure}

Figure \ref{magnets_MEFISTO} shows the arrangement of the permanent magnets in the MEFISTO ion source. The six permanent magnets of the ring-shaped cluster on the left are all magnetized inwards. Similarly the cluster on the right side is magnetized outwards. These two outer ring-shaped clusters form the magnetic field of an axial magnetic bottle and thereby establish the axial confinement of the plasma electrons. In contrast the central magnet cluster consists of permanent magnets which are magnetized alternatively inwards and outwards. This central cluster forms the hexapole field which produces the radial confinement of the plasma electrons. The ECR plasma is located inside this central ring arrangement. In addition Figure \ref{magnets_MEFISTO} also shows the helix antenna which radiates the microwave power into the ECR plasma.

The ionization process inside an ECR ion source works by electron collision with neutrals and ions. The electrons are effectively heated by the microwaves, most effectively for those meeting the electron cyclotron resonance condition. Equation \ref{ECR_omega} shows that the ECR condition in general and the resonance frequency $\omega$ in particular depend on the local magnetic field density $B$, the electron mass $m_{\rm{e}}$ and the charge of the electron $e$. However, the resonance frequency $\omega$ does not depend on the electron energy as long as the electron is not accelerated close to the speed of light where it's change of mass would come into account due to special relativity.

\begin{align}
\omega_{\rm{ECR}} = \frac{eB}{m_{\rm{e}}}
\label{ECR_omega}
\end{align}

Therefore the choice of the excitation frequency band determines the required magnetic field density $B$. Hence the design parameters of the magnet structure have to be chosen carefully so that the magnetic field fulfills the ECR criteria (\ref{ECR_omega}). This criteria is generally fulfilled on a closed iso surface of the magnetic field density (see Section \ref{The ECR zone}). This iso surface defined by the magnetic field distribution and the microwave frequency has to be located entirely inside the plasma container for optimal ECR efficiency.

This ECR ion source is used for the calibration of space craft instruments which are designed to detect highly charged ions in space. The performance of the ECR ion source with respect to the production of these highly charged ions depends on the square of the ECR resonance frequency \cite{Geller}. In addition the magnetic field distribution forms a magnetic mirror with a ratio $r_{\rm{m}} = B / B_{\rm{centre}}$ in all directions, which should be as high as possible for optimal electron and ion confinement. However the successful extraction of highly charged ions requires a locally poor confinement performance so that the ions can be extracted from the loss cone in the axial beam direction. In Figure \ref{magnets_MEFISTO} this extraction system is formed by the two cylinder shapes to the right.

All of these requirements have to be met for the successful operation of an ECR ion source. To achieve these goals three dimensional numerical simulations of the magnetic field can be used which allow the highly accurate design of future ECR ion sources. The simulation technology has been tested and the results have been compared with measurements of the existing MEFISTO ECR ion source.

Due to the unique field distribution of hexapole confined ECR ion sources the high energy electron population features a trident shaped spatial distribution as predicted theoretically by numerous authors (\cite{Chen general}, \cite{Geller}) and verified in experiments by Zschornack et al. (\cite{Zschornack}). The presented magnetic field simulation confirms both theory and experiment in three dimensions.

\section{The finite element model}

To achieve the highest possible simulation precision given the available infrastructure a finite element solver has been chosen. Magnetic simulations have been performed using \textit{Tosca} by \textit{Vector Fields} (www.vectorfields.com). All pieces and volumes relevant for the simulation of the stationary magnetic field were modeled carefully. The resulting finite element mesh itself has been tested extensively with reference patterns and circular charged particle trajectories such as electrons and ions. No deviation from the circular orbits were discovered within the system limit of 5000 integration steps for each particle. A summary of the FEM mesh characteristics is given in Table \ref{FEM_summary}.

\begin{table}[h!]
\centering
\begin{tabular}{lr}
number of linear elements & $2.4*10^6$ \\
number of quadratic elements & $1.9*10^5$ \\
number of nodes & $7.2*10^5$ \\
number of edges & $3*10^5$ \\
RMS & $1.033$  \\
distortion of worst element & $8.31*10^{-4}$
\end{tabular}
\caption{\textit{Some characteristics of the finite element model.}}
\label{FEM_summary}
\end{table}

\begin{figure}[h!]
\includegraphics[angle=0, width=0.5\textwidth]{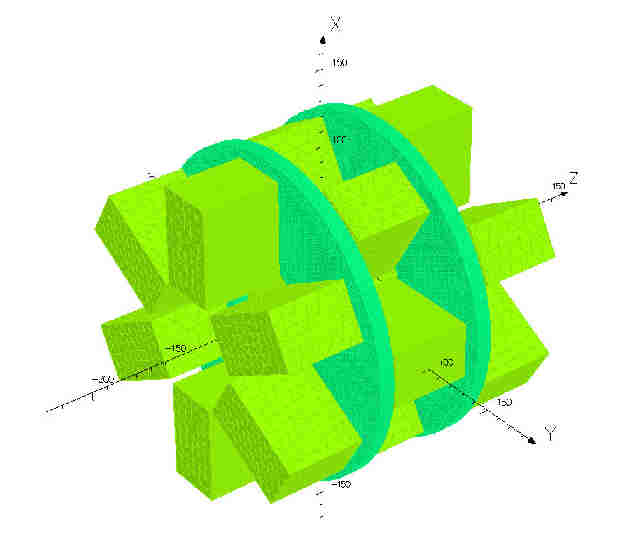}
\caption{\textit{The finite element model with its essential parts.}}
\label{FEM_model}
\end{figure}

Figure \ref{FEM_model} shows the finite element model used to simulate the magnetic field. 

\begin{figure}[h!]
\includegraphics[angle=0, width=0.5\textwidth]{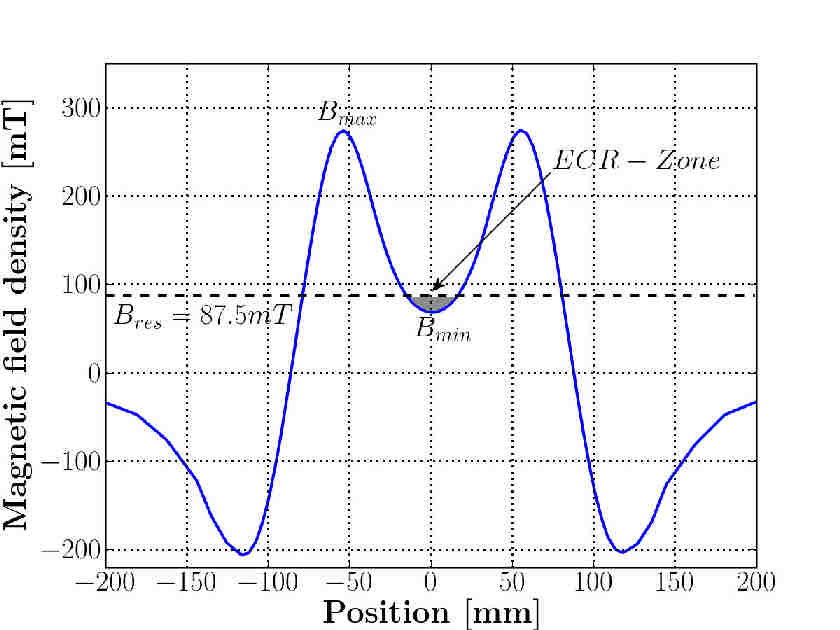}
\caption{\textit{Simulated magnetic field density along the beam axis.}}
\label{FEM_Bz}
\end{figure}

As described in Section \ref{intro} the model consists of three sets of block magnets arranged in rings around the central plasma area and two massive soft iron rings separating each of the block magnet ring arrangements. Figure \ref{FEM_Bz} shows the magnetic field density along the beam axis with respect to the z-coordinate. The ECR zone is indicated at $87.5~\rm{mT}$ where the field density meets the ECR condition.

The axial magnetic field of the existing MEFISTO ECR ion source has been measured previously \cite{AdrianMarti} \cite{Liehr_Trassl}. The local maxima and the local minimum of the measured magnetic field is given with 240 mT and 60 mT respectively. The corresponding mirror ratio is $r_{\rm{m}} = 240~\rm{mT}/60~\rm{mT} = 4$. The maxima of the numerical simulation are 240.1 mT, the minimum is 60.05 mT and the mirror ratio $r_{\rm{m}} = B_{\rm{max}} / B_{\rm{min}}$ is 3.998. The numerical model is therefore in excellent agreement with the measurement.  All subsequent simulations and results are based on the results of the presented numerical field simulation.

\section{The ECR zone}
\label{The ECR zone}

\begin{figure}[h!]
\includegraphics[angle=0, width=0.5\textwidth]{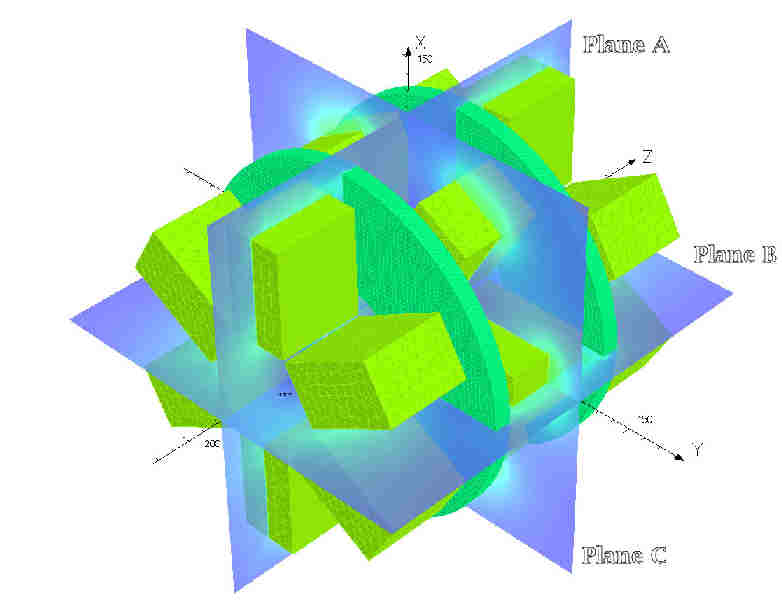}
\caption{\textit{To map the magnetic field density onto two dimensional patches we introduce cut planes in the 3D model.}}
\label{ECR_cut_faces}
\end{figure}

The ECR ion source is fed with a constant microwave frequency of 2.45 GHz, so the location of the resonance region can be determined by the corresponding magnetic field density $B_{\rm{res}} = 87.5~\rm{mT}$ (\ref{ECR_omega}). To present a two dimensional analysis of the field distribution and the ECR locality inside the magnet system we introduce cut planes (Plane A, B, C) into the three dimensional model as shown in Figure \ref{ECR_cut_faces}.

\begin{figure}[h!]
\includegraphics[angle=0, width=0.5\textwidth]{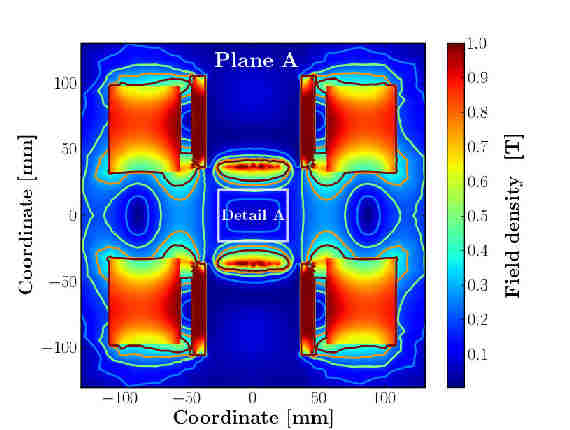}
\caption{\textit{Simulated magnetic field density in plane A.}}
\label{contour_A}

\includegraphics[angle=0, width=0.5\textwidth]{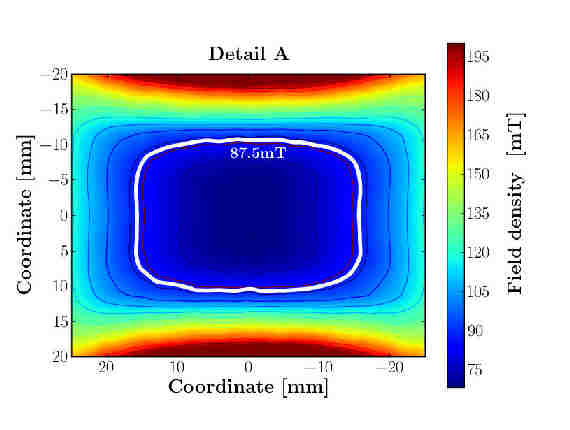}
\caption{\textit{Detailed view of the center part of plane A. The ECR iso contour line is highlighted. Due to the orientation of the cut plane (see Fig. \ref{ECR_cut_faces}) the shape of the iso contour line is almost symmetric.}}
\label{contour_A_detail}
\end{figure}

\begin{figure}[h!]
\includegraphics[angle=0, width=0.5\textwidth]{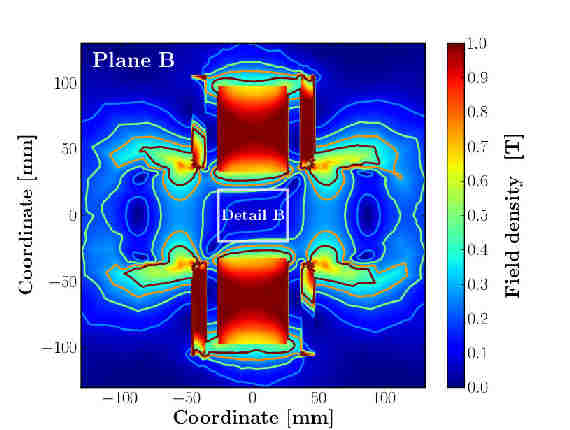}
\caption{\textit{Simulated magnetic field density in plane B.}}
\label{contour_B}
\includegraphics[angle=0, width=0.5\textwidth]{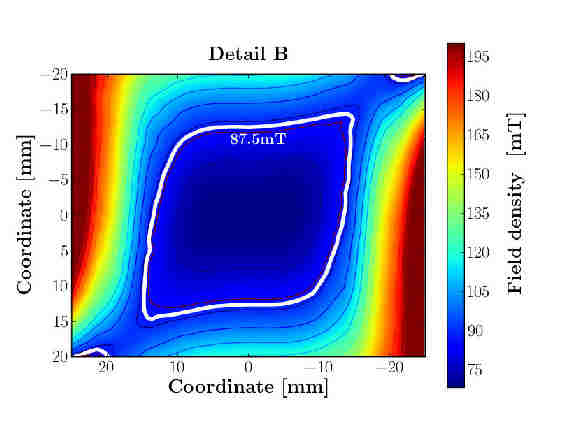}
\caption{\textit{Detailed view of the center part of plane B. The ECR iso contour line is highlighted. Cut plane B is rotated by 90 degrees with respect to plane A. The contour lines are no longer symmetric with respect to the beam axis.}}
\label{contour_B_detail}
\end{figure}

\begin{figure}[h!]
\includegraphics[angle=0, width=0.5\textwidth]{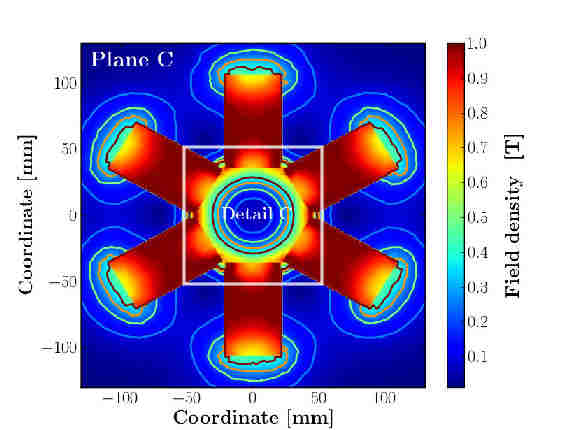}
\caption{\textit{Simulated magnetic field density in plane C.}}
\label{contour_C}
\includegraphics[angle=0, width=0.5\textwidth]{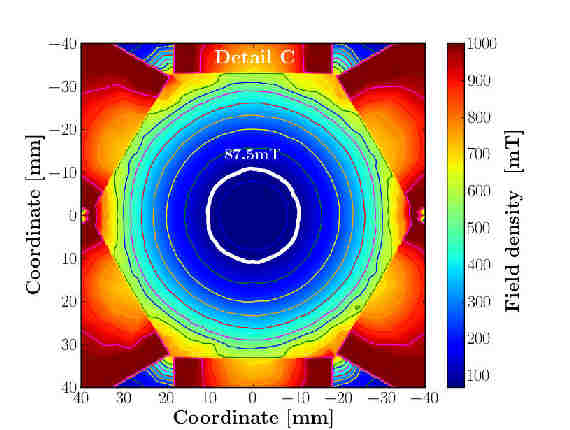}
\caption{\textit{Detailed view of the center part of plane C. The ECR iso contour line is highlighted. Toward the center the iso contour lines are concentric circles.}}
\label{contour_C_detail}
\end{figure}

In the following figures the magnetic field density has been mapped to color space. There are two figures of each cut plane (see Fig. \ref{contour_A} to \ref{contour_C_detail}) giving an overview and a detailed view of the central part of each plane respectively. Iso contour lines of the field density are also indicated on each graph.

Because cut plane B is rotated by 90 degrees with respect to cut plane A the magnetic field is cut at a different angle too revealing its non symmetric distribution in cut plane B (Figure \ref{contour_B_detail}). This feature will be visible more clearly in the 3D view presented further down this Section.

Despite the hexapole arrangement of the central magnet cluster the resulting field distribution in the symmetry plane C (see Fig. \ref{ECR_cut_faces}) yealds concentric iso contour lines as presented in Figures \ref{contour_C_detail}. However this is true only for contour lines which are close to the center of the hexapole. Contour lines further from the center and closer to the actual block magnets show the expected hexapole pattern with its distinct poles.

The small irregularities within the contour lines are due to the discretization of the model by the finite element size. The resulting relative error in the field density is lower than $10^{-2}$ and has been chosen to be acceptable.

\begin{figure}[h!]
\includegraphics[angle=0, width=0.5\textwidth]{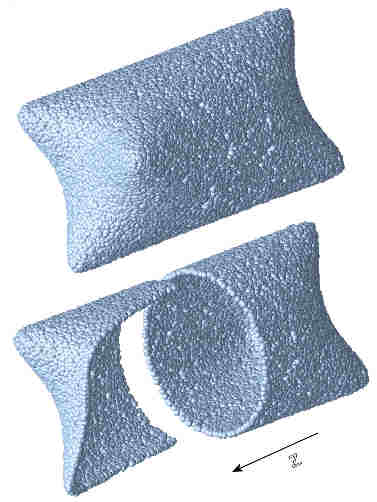}
\caption{\textit{Isosurface of a constant magnetic field density $B_{res}$ (upper image) and cut-away}}
\label{shape}
\end{figure}

In each detailed view above the iso contour line of 87.5mT is highlighted. Extending the iso contour line concept from two to three dimensions leads to an iso contour surface. This iso contour surface includes all element centroids of the finite element model which fulfill the criteria of the given magnetic field density $B_{ref} = 87.5~\rm{mT}$. On this surface the gyration frequency of the electrons in the magnetic field equals the microwave frequency. The plasma electrons are therefore heated effectively on this surface. The surface defined by this criteria is shown in Figure \ref{shape}. 

On Figure \ref{shape}, the centroid of each FEM element of the specified surface is represented by a small bubble. Each bubble also represents one launch coordinate for the trajectory model represented in the following section. The shape is symmetric with respect to the origin located at the center of the shape. Note the trident shapes on both ends. These tridents can be explained as follows. Because the locations of the trident shapes have an offset from the center of the magnet arrangement in positive and in negative z direction the radial magnetic field component of the outer ring magnet clusters no longer cancel out. This leads to a dominance of the outwards magnetized cluster in positive z direction. Consequently the radial field of the outward magnetized hexapole magnets is enhanced and the field of the inwards magnetized hexapole magnets is suppressed where the trident shape in positive z direction is located. The trident shape in negative z direction is formed in the same way with the field of the inward magnetized hexapole magnets suppressed and the field of the outwards magnetized hexapole magnets enhanced. Hence, the shapes are twisted relative to each other by 60 degrees because of the same angle separating inward and outward magnetized permanent magnets in the hexapole cluster. The cross section through the mid plane is circular (see Fig. \ref{contour_C_detail} for 2D contour plot) and has a diameter of 21.0 mm. The overall length of the shape is 37.3 mm.

\begin{figure}[h!]
\includegraphics[angle=0, width=0.5\textwidth]{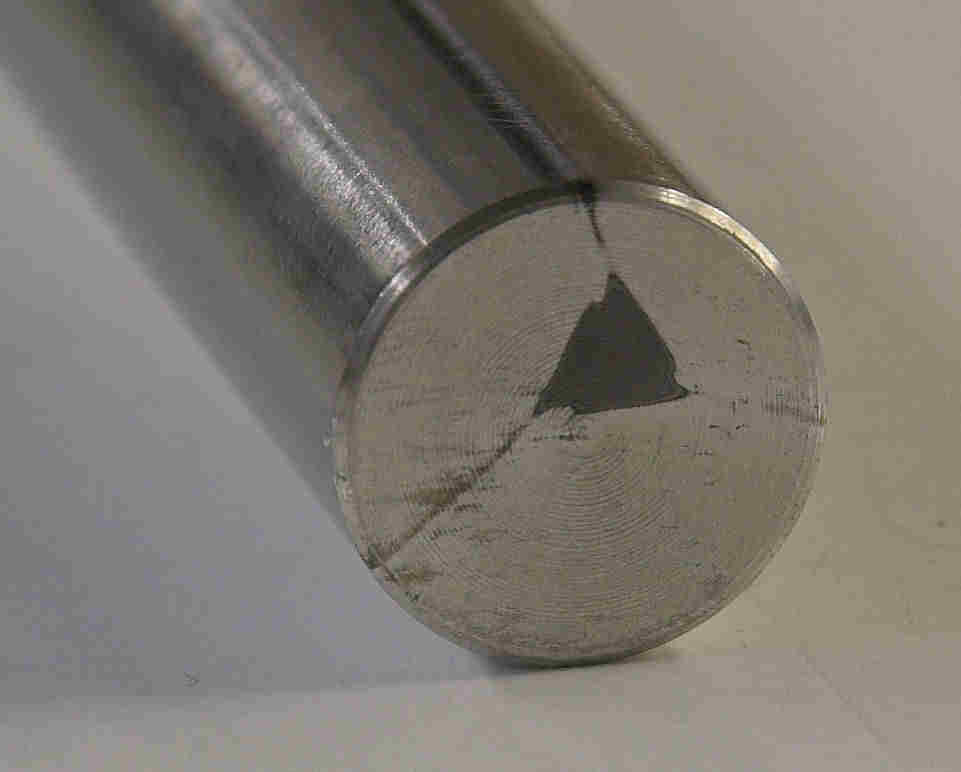}
\caption{\textit{The ion beam of hexapole confined ECR ion sources produces trident shaped sputtering patterns (Picture from CASYMS ECR ion source, University of Bern). Here two trident patterns are superimposed on a microwave antenna. This antenna was used to radiate the microwaves into the plasma and for impedance matching. It was in contact with the ECR zone.}}
\label{photo_trident}
\end{figure}

The described trident shapes of the ECR zone also influence the plasma as explained further down. Figure \ref{photo_trident} shows a photograph of two superimposed sputtering tridents observed in a hexapole confined ECR facility.

\begin{figure}[h!]
\includegraphics[angle=0, width=0.5\textwidth]{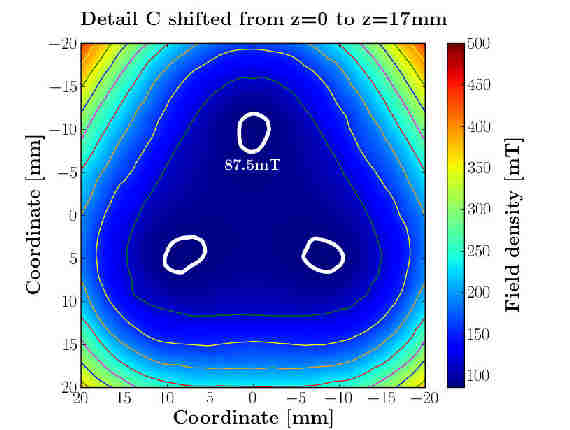}
\caption{\textit{Simulated magnetic field density in a plane parallel to C, but shifted by 17mm. While contour plot C at z=0 (Fig. \ref{contour_C_detail}) features circular iso contour lines another contour plot at z=17mm reveals the similarity to the sputtering trident. The three separated islands of the ECR iso contour lines are highlighted.}}
\label{contour_C_detail_shifted}
\end{figure}

Cut plane C is located at z=0 which results in circular iso contour lines of the field density near the center not similar to the observed trident shaped sputtering patterns. However if the cut plane is shifted parallel to z=17mm the iso contour lines be identified with the sputtering trident (see Fig. \ref{contour_C_detail_shifted}). This is because the shifted contour plane cuts the isosurface (Fig. \ref{shape}) no longer at the center (as plane C is doing it) but 17mm toward the trident face.

We can clearly identify the top face of the shape with the typical sputtering trident well known for ECR ion sources with a hexapole radial confinement such as MEFISTO. The identification of the simulated and the observed trident takes into account the shape of the ECR region, the presence of high energetic electrons and the subsequent emergence of ions as follows. 

According to the successful ECR plasma model of Wurz et al. (\cite{Wurz}) the ionization is a step by step process starting with few free electrons and a majority of neutrals. The few free electrons get accelerated to high energies by the electric field of the microwave injection in the described way of electron cyclotron resonance. The energetic electrons then collide with neutrals to produce single charged ions and more free electrons. More neutrals get ionized due to the higher electron density. Simultaneously single charged ions collide with energetic electrons to be ionized from single to double charged ions and so forth. This way very few free electrons create an avalanche effect and finally establish the quasi stable electron density of the ECR plasma featuring highly charged ions. A fully consistent version of this model has been tested numerically and verified up to a charge state of Ar8+ by Hohl et al. (\cite{Hohl2}).

The magnetic trident face has been identified as part of the volume in which ions are created. Ions are also created by the same process in the rest of the depicted shape (Fig. \ref{shape}) as well as inside of it. However only the trident part of the front face is being mapped by the extraction ion optics to the target surfaces where the sputtering has been observed. 

The magnetic field density increases in every direction pointing away from the entire depicted shape. The closed depicted shape therefore confines a volume of minimal magnetic flux density. The magnetic arrangement thereby fulfills the criteria of a minimum B-field structure. Further we have the liberty to refer to the defined volume as a plasmoid - a plasma magnetic entity \cite{Bostick}. The presented isosurface element centroids were used to define the launch points of the electron trajectories described in the next section.

\section{The trajectory model}

The ECR plasma is charged slightly negative as described by Shirkov \cite{shirkov_1}, \cite{shirkov_2} and verified experimentally by Golovanivsky and Melin \cite{Golo}. According to these measurements the plasma potential affecting the electron trajectories has been chosen to be minus 10 V. Due to the compact shape of the simulated plasmoid the spatial distribution of the plasma potential was modeled as a sphere with the same diameter as the plasmoid.

The electron heating takes place within the ECR qualified volume sheath with a magnetic field density of $87.5~\rm{mT}$ including an assumed tolerance of $\pm 0.5~\rm{mT}$, which is due to the bandwidth of the implemented microwave generator. Because of collisions and momentum transfer the production of new ions can only happen close to where the electrons are energized. Consequently the initial launch coordinates of the simulated electrons were chosen to be located within the ECR qualified volume sheath. Each bubble seen in Figure \ref{shape} is located at an element centroid of the magnetic FEM model and therefore represents one launch point for the trajectory model. In total the iso surface shape yields 25211 launch points.

\begin{figure}[h!]
\includegraphics[angle=0, width=0.5\textwidth]{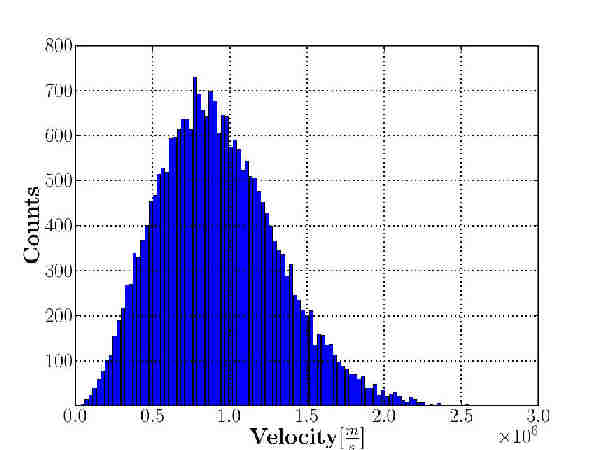}
\caption{\textit{The velocity component parallel to the local magnetic field is distributed with a corresponding temperature of 2eV.}}
\label{isotropical_vel_distr}
\end{figure}

The ECR heating process mainly takes place perpendicular to the local magnetic field lines \cite{Geller}. Therefore the initial velocity distribution has been chosen to be anisotropic. The velocity component parallel to the local magnetic field of each trajectory was scattered symmetrically around zero with a maxwellian temperature of 2eV \cite{Hohl}. Figure \ref{isotropical_vel_distr} shows a histogram of the modulus of this vector component.

\begin{figure}[h!]
\includegraphics[angle=0, width=0.5\textwidth]{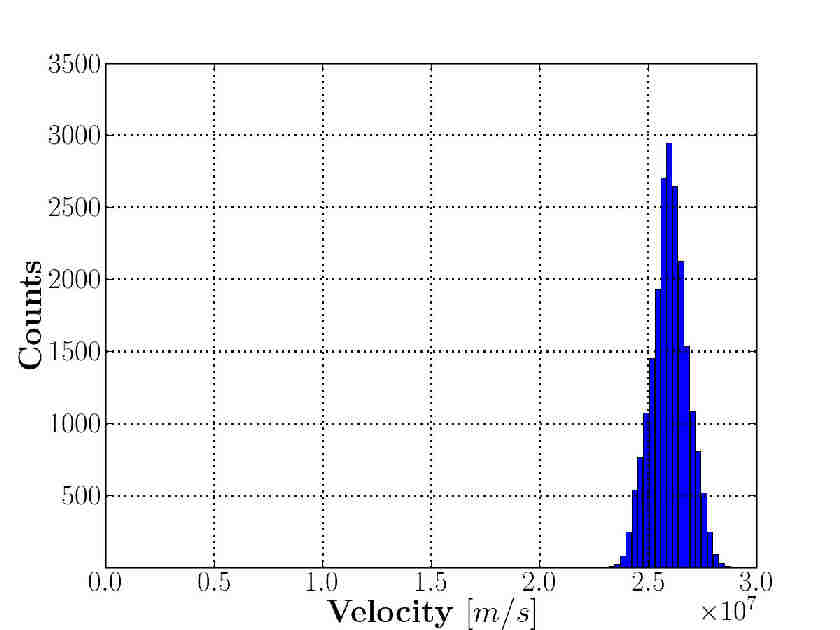}
\caption{\textit{The norm of the velocity component perpendicular to the magnetic field has an offset of 2keV and a corresponding temperature of 2eV with respect to the offset.}}
\label{anisotropical_vel_distr}
\end{figure}

In contrast the modulus of the two velocity components perpendicular to the magnetic field were scattered around a kinetic particle energy of 2keV corresponding to the measurements of M. Hohl et al. \cite{Hohl}. Figure \ref{anisotropical_vel_distr} shows a histogram of the modulus of the perpendicular velocity components centered around 2keV with a maxwellian temperature of 2eV with respect to the offset.

The model does not take into account any particle scattering phenomena nor any other kind of particle particle interaction. The element size in the finite element and the selection criteria of the ECR zone gave a total number of launched particles of 25211. All simulations have been performed on a IntelCore2Duo 6700 clocked at 2.66GHz and 2 GB of RAM. The average run time was 83 hours at more than 99 $\%$ of CPU load.

\section{Results}

The trajectory lifetimes, lengths and velocities of each particle were analyzed. Due to hardware limitations the simulations were stopped once less than one percent of the initial number of electrons remained in the simulation. The life time of a particle is limited by the collision with a wall of the plasma container or by the end of the simulation. In a histogram of electron trajectory counts versus trajectory life time we can fit two exponential decay functions of the electron life times $\tau_{\rm{1}}$ and $\tau_{\rm{2}}$:

\begin{align}
n(t) = n_0 \left(e^{- \frac{t}{\tau_1}ln(1/2)} + \frac{1}{k}e^{- \frac{t}{\tau_{\rm{2}}}ln(1/2))}\right)
\label{Exp_decay}
\end{align}

\begin{figure}[h!]
\includegraphics[angle=0, width=0.53\textwidth]{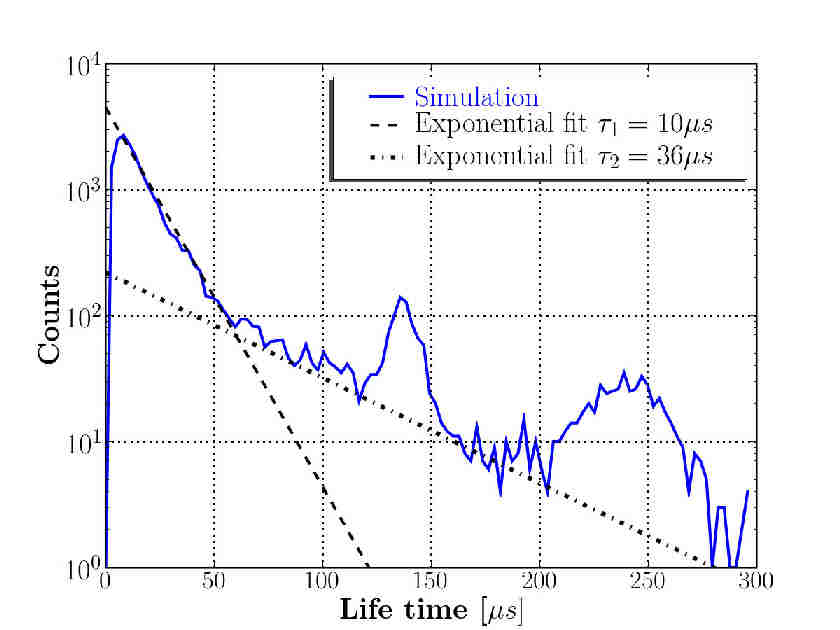}
\caption{\textit{Histogram of particle life time and the corresponding exponential fits.}}
\label{hist_time}
\end{figure}

Figure \ref{hist_time} shows a histogram of the particle life time and an exponential decay with $\tau_1 = 10 \mu s$ and another one with $\tau_2 = 36 \mu s$ indicated.

\begin{figure}[h!]
\includegraphics[angle=0, width=0.53\textwidth]{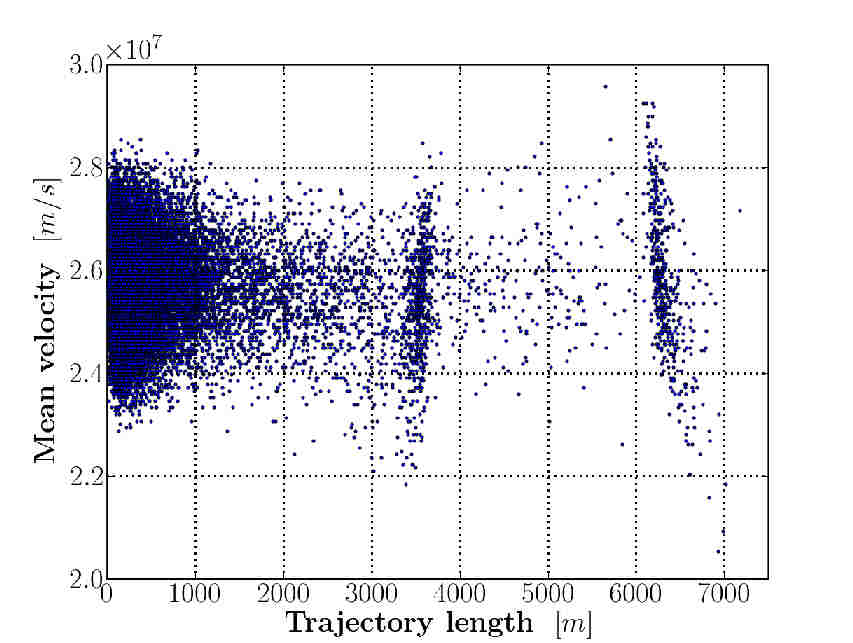}
\caption{\textit{Mean velocity versus trajectory length of simulation A. Two clusters at 3537m and 6287m are separated from the main population}}
\label{length_vel}
\end{figure}

There are two significant deviations from the exponential decay behavior around $136 \mu s$ and lesser so at $242 \mu s$. The deviation represents two electron fractions featuring unexpected long life times. This simulation will be referred to as simulation A. Figure \ref{length_vel} gives the mean velocity of each trajectory with respect to its length for this simulation.

Apart from the main population from 0 to 2000m there are two clusters of trajectories accumulating at a trajectory length of 3537m and another one at 6287m. Both cluster areas present a wide velocity spectrum centered around the same value as the main population. The corresponding trajectories have been traced back to find their underlying launch points. The respective launch points and their initial conditions are distributed both homogeneous over the plasmoid shape as well as correctly anisotropic in velocity space. Hence, the phenomena is not the result of a special launch location of the trajectories nor is it due to an anomaly of the initial velocities.

\begin{figure}[h!]
\includegraphics[angle=0, width=0.53\textwidth]{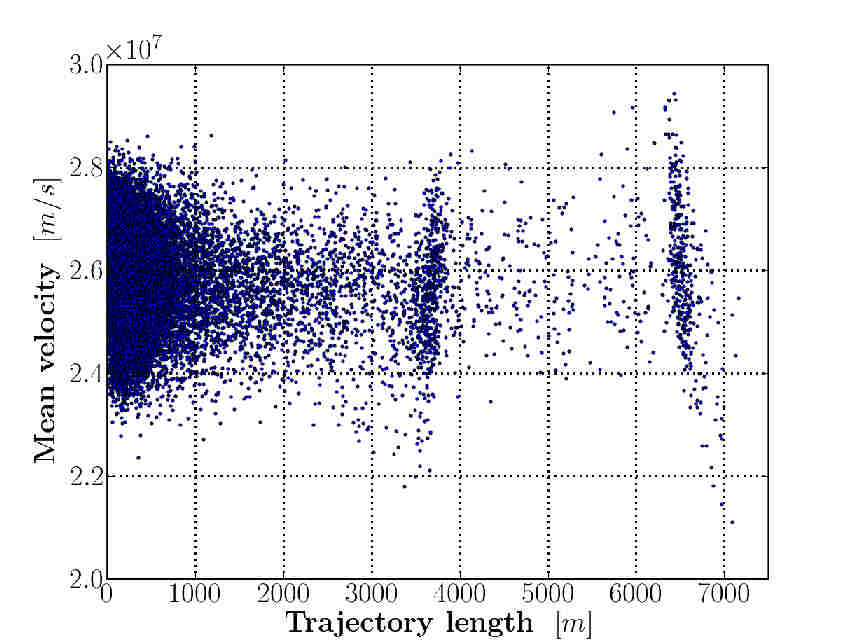}
\caption{\textit{Trajectory length versus mean velocity of electrons in second simulation run B.}}
\label{length_vel_second}
\end{figure}

To distinguish the simulated phenomena from numerical errors associated with the random particle velocity distribution a second simulation was performed. The second simulation made use of the same principle of anisotropic initial velocity distribution as the first simulation but starts with a new set of particles. Due to the randomized velocity vector creation process the new set is deliberately different from the first one despite the equal mean energy and the equal initial start locations. Figure \ref{length_vel_second} shows the mean velocity versus trajectory length of the second simulation B.

Table \ref{stats_vel_length} gives a statistical breakdown of the trajectory lengths of both simulations.

\begin{table}[h!]
\centering
\begin{tabular}{lll}
Simulation & $\mu$ & $\sigma$  \\
\hline
First simulation A & $296.1m$ & $1207.6m$ \\
Second simulation B & $343.8m$ & $1213.5m$ \\
\end{tabular}
\caption{\textit{A statistical breakdown of the trajectory length of both simulations. $\mu$ refers to the median and $\sigma$ to the standard deviation.}}
\label{stats_vel_length}
\end{table}

The high standard deviation indicates the wide spread of the length distribution. The medians of both simulations show a shift of 47.7m. We can see the graphs for both simulation runs are similar in overall shape. Both show the same two clusters in the same area. However in the second simulation the clusters are slightly shifted toward a higher trajectory length which corresponds to the higher median of the second simulation. Table \ref{sum_vel_length} summarizes the findings related to both clusters.

\begin{table}[h!]
\centering
\begin{tabular}{lll}
Simulation & $\mu_{cluster 1}$ & $\mu_{cluster 2}$ \\
\hline
First simulation A & $3537m$ & $6287m$ \\
Second simulation B & $3631m$ & $6499m$ \\
$ \Delta_{rel}$ & $2.59 \%$ & $3.26 \%$ \\ 
\end{tabular}
\caption{\textit{A summary of the clusters separated from the main population. $\mu_{cluster 1}$ refers to as the median of the first cluster in each simulation and $\mu_{cluster 2}$ to the second one. $\Delta{rel}$ depicts the relative difference between each median.}}
\label{sum_vel_length}
\end{table}

Due to the different initial conditions of velocity distribution the two simulation runs are not identical. Both clusters show a relative shift between $2.59 \%$ and $3.26 \%$. However both simulation runs show the clusters are clearly separated from the main population and grouped in the same range of trajectory length. Neither median of either simulation corresponds to a dimension given by the plasma container nor a low multiple integer of such. 

\section{Discussion}

Moving electrons in a magnetic field are deviated from straight trajectories by the Lorentz force perpendicular to their flight directions. The electrons enter circular orbits with a radius $r_L$. This is called the Larmour radius of an electron. It depends solely on the electrons energy and the local magnetic field. It is given by:

\begin{align}
r_L = \frac{vm}{eB}
\label{larmour}
\end{align}

where $v$ is the velocity of the electron, $m$ its mass, $e$ its charge and $B$ is the local magnetic field density. For the case of the MEFISTO ECR zone and 2keV electrons this results in $r_L = 1.722mm$. The Larmor radii in the ECR zone are therefore much smaller than the plasma container. Hence we can give a rough approximation of revolutions an electrons takes before colliding with a container wall with $\mu$ as the median of the first simulation run (\ref{revs}). 

\begin{align}
n_{rev} \cong \mu / (2 \pi r_L) = 2.7*10^4
\label{revs}
\end{align}

Given a collisionless plasma the electron motion in the simulated magnetic arrangement is therefore dominated by the magnetic field rather than the container dimensions and could indeed produce long trajectories as it is suggested by the simulation results. However this does not apply for the current operation parameters of the MEFISTO ion source as we will see in \textit{Section \ref{section_Relevance}}.

\subsection{Relevance to laboratory plasmas}
\label{section_Relevance}

Laboratory plasmas as in the MEFISTO facility are not collisionless. The mean free path is limited not only by the plasma container walls but also by collisions with other particles of the plasma. According to Chen \cite{Chen general} the mean free path of an electron is given by (\ref{formula_mean_free_path}):

\begin{align}
\lambda = 3.4*10^{11} \frac{T_{eV}^2}{n \ln \Lambda} \left[ m \right]
\label{formula_mean_free_path}
\end{align}

\begin{align}
\Lambda = 12 \pi n \lambda_D^3
\label{formula_mean_free_path_1}
\end{align}

$\lambda_D$ is the Debye length given by:

\begin{align}
\lambda_D = \sqrt{\frac{\epsilon_0 K T_e}{n  e^2}}
\label{formula_mean_free_path_2}
\end{align}

where $T_{eV}$ is the electron temperature, K is the Boltzmann constant, n is the plasma density and e is the charge of the electron. Assuming a plasma density of  $10^{16}$ $ m^{-3}$ and an electron temperature of 2keV we get a Debye length $\lambda_D = 3.24mm$, a $\Lambda$ of $1.278*10^{10}$ and a mean free path length $\lambda$ of $5.84m$.

This value clearly shows the mean free path length of 2keV electrons in an ECR laboratory plasma are far too short to produce the phenomena suggested by the trajectory simulation. However a plasma density of $10^{13}$ $ m^{-3}$ would lead to a mean free path of $5089m$ by the same calculation. This would bring the simulation results into physical possibility. 

If we assume an effective cross section for electron collisions with neutrals of $10^{-20}$ $m^2$ \cite{Geller} and again a plasma density of $10^{16}$ $ m^{-3}$ we get a mean free path to the next neutral particle of $\lambda_n = 10^4 m$. Collisions with neutrals are therefore not limiting.

\subsection{Relativistic considerations}

Due to the electron energies involved of 2keV a slight change of electron mass occurs due to special relativity. The electron cyclotron resonance condition is attached to the electron mass and the magnetic field density (\ref{ECR_omega}). A change in electron mass therefore results in a different resonance condition of the qualified magnetic field density. The relation is as follows:

\begin{align}
\frac{B_1}{B_0} = \frac{m_1}{m_0} = \gamma =\frac{1}{\sqrt{1-(\frac{v}{c})^2}}
\label{formula_relativity_1}
\end{align}

Where $v$ is the velocity of the electron, $c$ the speed of light in vacuum, $B_1$ the ECR qualified magnetic field density with the consideration of mass increase and $B_0$ without the consideration of mass increase. $v / c$ equals $8.67 * 10^{-2}$ and  $B_1 / B_0 -1$ equals $3.78 * 10^{-3}$. This modified magnetic field requirement would be fulfilled at a different location due to the spatial distribution of the magnetic field. In fact the isosurface depicted in \textit{Section 3} would be enlarged toward the plasma chamber walls where the magnetic field features a higher modulus and where the resonance condition would again be fulfilled. However the shift of the isosurface would be smaller than the thickness of the surface due to the implemented frequency tolerance of the microwave generator. Hence this drift in location of the electron cyclotron resonance toward a higher magnetic field locality due to the effect of special relativity has been neglected.

\section{Conclusions}

The plasmoid shape resulting from the magnetic configuration of MEFISTO is responsible for the typical sputtering geometry of hexapole confined ECR ion sources. Future permanent magnet ECR ion sources can be designed very accurately using finite element modeling of the magnetic field. The accuracy of the FEM model was confirmed by comparing it to experimental data on MEFISTO. The location of the ECR qualified plasmoid constrained by an isosurface of constant magnetic field density depends on the magnetic arrangement only. It can therefore be chosen arbitrarily by tuning the design parameters of the magnetic setup.

Confined by permanent magnet hexapoles, long electron life times can be expected for low density plasmas in ECR ion source. This could also be very useful for electron or positron traps or any other particle storage facility as the numerical model is scalable by particle mass, particle energy and magnetic field density.

\end{document}